\documentclass[useAMS,usenatbib]{mn2e}
\usepackage{graphicx}
\usepackage{rotating}
\usepackage{longtable}
\usepackage{lscape}
\usepackage{amssymb,amsmath}
\begin{document}

\title[Low z QSO properties]
{Low redshift  quasars in the SDSS Stripe 82. Host galaxy colors and close environment }

\author[Bettoni  et al.]{
	D. Bettoni$^{1}$\thanks{E--mail: {\tt daniela.bettoni@oapd.inaf.it}}, 
        R. Falomo$^{1}$,
	J.~K.~Kotilainen$^{2}$, 
	K. ~Karhunen$^{3}$, and 
	M. Uslenghi$^{4}$ \\
       	$^{1}$ INAF -- Osservatorio Astronomico di Padova, Vicolo dell'Osservatorio 5, I-35122 Padova (PD), Italy\\
       	$^{2}$ Finnish Centre for Astronomy with ESO (FINCA), University of Turku, V\"ais\"al\"antie 20, FI-21500 Piikki\"o, Finland\\
       	$^{3}$ Tuorla Observatory, Department of Physics and Astronomy, University of Turku, FI-21500 Piikkio, Finland.\\
       	$^{4}$ INAF-IASF via E. Bassini 15, 20133 - Milano, Italy
	}
\maketitle
\label{firstpage}

\begin{abstract}

We present a photometrical and morphological multicolor study of the properties of low redshift (z$<$0.3) quasar hosts based on a large and homogeneous dataset of quasars derived from the Sloan Digital Sky Survey (DR7). We used quasars that were imaged in the SDSS Stripe82 that is up to 2 mag deeper than standard Sloan images. This sample is part of a larger dataset of $\sim$400 quasars at $z<0.5$ for which both the host galaxies and their galaxy environments were studied \citep{falomo14,karhunen14}. For 52 quasars we undertake a study of the color of the host galaxies and of their close environments in $u,g,r,i$ and $z$ bands.  
We are able to resolve almost all the quasars in the sample in the filters $g,r,i$ and $z$ and also in $u$ for about 50\% of the targets. We found that the mean colors of the QSO host galaxy ($g-i$=0.82$\pm$0.26; $r-i$=0.26$\pm$0.16 and $u-g$=1.32$\pm$0.25)  are very similar to the values of  a sample of inactive galaxies 
matched in terms of redshift and galaxy luminosity with the quasar sample. There is a suggestion that the most massive QSO hosts have bluer colors.
Both quasar hosts and the comparison sample of inactive galaxies have candidates of close ($<$ 50 kpc) companion  galaxies for  $\sim$ 30\% of the sources with no significant difference between active and inactive galaxies.
We do not find significant correlation between the central black hole (BH) mass and the quasar host luminosity that appears to be extra luminous at a given BH  mass with respect to the local relation (M$_{BH}$ -- M$_{host}$) for inactive galaxies. This confirms previous suggestion that a substantial disc component, not correlated to the BH mass, is present in the galaxies hosting low z quasars. 
These results support a scenario where the activation of the nucleus has negligible effects on the global structural and photometrical properties of the hosting galaxies.

\end{abstract}

\begin{keywords}
{galaxies: evolution ---  galaxies: active --- galaxies: nuclei --- quasars: general}
\end{keywords}

\section{Introduction}

The characterization of the properties of the host galaxies of QSO is an important tool to investigate the role of the AGN in their evolution. In the last years there is growing evidence that \citep[see e.g.][]{Schawinski,heckman14} AGN host are very similar in morphology to inactive galaxies at same redshift. For instance \citet{cisternas11} found that the host galaxies have normal morphologies in $\sim$85\% of their sample of  X-ray bright AGN. 

The characterization of the properties of AGN hosts offers also the opportunity to investigate the link between the central black hole mass and its host galaxy at moderate to high redshift and to trace the possible co-evolution at different cosmic epochs.
The mass of the central black hole (BH) can be derived under the assumption that the broad emitting regions are under the sphere of influence of the supermassive BH using the virial method from the analysis of the broad emission lines of the QSO and from empirical relation between the continuum luminosity and the size of the Broad Line Region (BLR) \citep{dunlop03,gultekin09}. 
Because of the high luminosity of QSO and the prominent emission lines these can be done for a large number of sources using various emission lines \citep[e.g.][]{shen11}.

On the other hand the characterization of the properties of their host galaxies is more challenging because one has to decompose the 
starlight of the host galaxy from that of the nuclear emission. Since the nucleus is more luminous than the host galaxy this observation requires excellent seeing conditions for ground-based observations or images obtained with Hubble Space Telescope. 
In spite of these difficulties for several QSO it was possible to de-blend the nuclear and starlight contribution of quasars \citep[see e.g.][]{McRi,kotilainen04,falomo14} using ground based data or using  HST imaging for relatively low (z $<$ 1) redshift objects \citep[see e.g.][]{bahcall97,dunlop03,floyd04,kukula01,ridgway01,jahnke09,pagani}.
For a limited number of QSOs the use of 8-10m telescope under superb seeing conditions or 
with adaptive optics  imaging has allowed the study of quasar host at high redshift 
 \citep{kotilainen07,kotilainen09} and/or with adaptive optics  \citep{falomo08}
 These observations allowed to trace a a first view of the cosmic co-evolution 
 of SMBH and their host galaxies \citep[see e.g.][]{decarli2010,decarli2012,Sanghvi14}.


 A less explored issue is to assess the stellar population of the galaxies hosting active SMBH as compared with that of inactive galaxies.
 The understanding of the link between stellar population and growth and activation of a massive BH can in fact offer important clues to  the role of merging for fueling the central BH. Although galaxy interactions and merging  have been long assumed as a main drive of the AGN phenomenon \citep[e.g.][]{Schawinski,cisternas11,Kocevski}  there are AGN surveys that seems to indicate that interactions do not lead to enhancement of nuclear activity \citep{li08}. AGN activity in interacting galaxies is no different from that observed for non interacting galaxies.

The study of the colors of the host galaxy is also important to characterize its nature. It is well known that  the color-magnitude relation for  normal galaxies exhibits two sequences. A {\it red sequence}, populated by massive, bulge-dominated galaxies with older, passively evolving stellar populations, and  a {\it blue cloud}, populated by blue, star-forming galaxies of small and intermediate masses \citep[e.g.][]{baldry,weiner}. Past studies indicated that the AGN host galaxy lie in the so called {\it green valley} that is the transition region between red sequence and blue cloud \citep{Silverman,Treister} this result suggested that the AGN feedback can be responsible in regulating the star formation moving galaxies from the blue cloud to the red sequence.

The best way to investigate the signature of induced starburst in active galaxies is through the optical spectra of the host galaxies. However this can be pursued only for a limited number of sources because it requires very efficient spectroscopic capabilities and observations under excellent conditions. This technique therefore has been used with success only for a limited number of objects at relatively low redshift and with a low luminosity nuclei \citep[see e.g.][]{Nolan,Miller,Canalizo}. More recently, using Integral Field Units (IFU) spectrographs, \citet{Liu} and \citet{Husemann} studied samples of luminous unobscured (type 1) quasars providing the morphology, kinematics and the excitation structure of the extended narrow-line region to probe relationships with the black hole characteristics and the host galaxy.

An alternative approach to obtain clues of recent star formation in the host of quasars is to measure the colors of the host galaxies from the de-blending of the nucleus and host components in multi color images of quasars. Although this cannot use the 
more powerful spectroscopic diagnostic to envisage the underlying stellar population this approach can be adopted for a larger sample of targets provided that the
multicolor images be available. This was done in past for  19 quasars at z $<$ 0.2 by \cite{jahnke04} who find  mixed results and for a large number of $z<0.3$ BL Lac host galaxies by \citet{kotilainen04} and \citet{Hyv} who find bluer than normal hosts. 
Quasar host that are dominated by a disc component appear to have similar color to that of inactive galaxies while quasars that have hosts dominated by spheroidal component appear bluer than inactive galaxies. 


More recently \cite{matsuoka14}(hereafter M14) and \citet{matsuoka15} analyzed the stellar properties of galaxies hosting optically luminous, unobscured quasars at z$<$ 0.6 using Stripe82 images. They focused on the colors of the host galaxy and found that quasar hosts are very blue and almost absent on the red sequence with a marked different distribution from that of normal (inactive) galaxies. 

In this paper we aim to investigate the colors of a sample of  $\sim$ 50 low redshift z $<$ 0.3 QSO using multicolor images obtained by SDSS in the Stripe82 area. This is a stripe along the Celestial Equator in the
Southern Galactic Cap. It is 2.5$^{\circ}$ wide and covers -50$^{\circ}\leq$RA$\leq$+ 60$^{\circ}$, so its total area is 275 deg$^2$. Stripe 82 was imaged by the SDSS multiple times since 2000 only under optimal seeing, sky brightness, and photometric conditions. The total number of images reaches $\sim$100 for the S strip and $\sim$ 80 for the N strip. The final frames were obtained by co-adding selected fields in r-band, with seeing (as derived from 2D gaussian fit of stars and provided by SDSS pipeline) better than 2", sky brightness $\leq$19.5 $mag/arcsec^2$ and less than 0.2 mag of extinction.The QSO sample is extracted from a much  larger ($\sim$ 400 objects) sample of low redshift quasars in Stripe82 for which we performed a complete study of the host galaxies in i-band and their large scale environments \citet{falomo14,karhunen14,karhunen15}.
Using the deep co-added SDSS images of Stripe82 we derive the properties in all five SDSS bands of the host galaxies and we compare with a control sample of (non-AGN), inactive galaxies. 

The paper is organized as follows: in Section 2 we present our QSO sample. Section 3 describes the analysis of the data and the main properties of the host galaxies and in Section 4 we discuss our results and we compare our findings with those of M14 and other previous studies. We adopt the concordance cosmology with H$_0$ = 70 km s$^{-1}$ Mpc$^{-1}$, $\Omega_m$ = 0.3 and $\Omega_\Lambda$ = 0.7.

\section{The low z QSO sample}

In  previous papers of this series  we   investigated the properties of the host galaxies \citep{falomo14}(hereafter F14) and of the galaxy environments \citep{karhunen14}  of a large ($\sim$ 400) dataset of low redshift ($z<0.5$) quasars extracted from the the fifth release of the SDSS Quasar Catalog  \citep{schneider2010}, based on the SDSS-DR7 data release \citep{abazajian09} and observed in the region of sky covered by the Stripe82 \citep{annis2011}. These studies were based on the images in $i$ band and allowed us to go about $\sim$2  magnitudes deeper with respect to the usual Sloan data and make possible the study of the QSO hosts and their environments. 

For this multicolor study we considered only objects at z $<$ 0.3 because beyond this limit the characterization of the QSO  host galaxies becomes arduous at bluer filters due to the reduced contrast between the host galaxy and the nuclear emission. We did a number of tests to evaluate the possibility to detect and measure reliably the quasar hosts for a significant fraction of the targets at various redshift and found that the best compromise that maximize the number of objects resolved in all filters (but not for all objects in $u$ band) is to set a redshift limit to z $\sim$ 0.3. In fact while in the $i$ filter the fraction of resolved objects can be as high as 70-80\% up to z $\sim$ 0.5 (cfr. F14; M14) this fraction falls below 50\% for filters $g,r$  and $\sim$  10\% in the $u$ band.

From our previous  sample of 416 QSO (F14), that were extracted from the QSO catalogue  \citep{schneider2010}, and imaged in the S82 region we extracted those with $z<0.3$. This yields 60 QSO, however 4 are unresolved also in our previous (F14) analysis and are not considered. One object has been removed from the original list due to the presence of a defect in the image. We visually inspected the spectra of the remaining 55 QSO and found that three objects have emission lines typical of LINERS ($H_{\beta}$ FWHM $<$ 1000 km/sec and log([OIII]$\lambda$5007/$H_{\beta})<0$, as derived from SDSS SpecLine table) \footnote{objects \#56~(SDSSJ214817.43+000419.8),\\ \#112~(SDSSJ225757.22+002608.3) and \\ \#172~(SDSSJ000834.71+003156.1) in F14}  and were eliminated from the sample. Under these assumptions we are able to construct a sample of 52 QSO, representing the 87\% of all QSO in Stripe82 and with  z$<$0.3, for which we can study the color of the host galaxy. 
 The mean redshift of this sub sample is $<z>$ =$0.25\pm0.06$ and the average absolute magnitude of quasar host $<M_i>$ = $-22.57\pm0.65$. The mean luminosity of our low z QSO is $M_B$=-21.8 as derived from the g apparent magnitude and transformed to B band using \citet{Jordi} transformations.
As expected the  sample is dominated by low luminosity QSO and merge into the region of objects that can be classified also as Seyfert 1 galaxies (classical division between QSO and Seyfert 1 \citep{SG} is $M_B$=-22.2 in our cosmology).
 
\begin{figure}
\centering
\includegraphics[width=0.9\columnwidth]{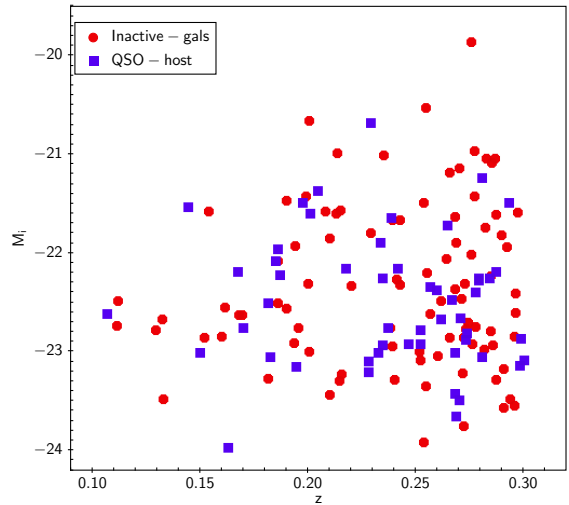}
\caption{The QSO hosts (blue squares) are compared to inactive galaxies (red circles) in the z-$M_i$ plane (see text).}
\label{fig:miz}
\end{figure}

In our previous work on the environment of low z QSOs \citep{karhunen14} we defined a control sample of galaxies with similar redshift and host galaxy absolute magnitude distributions. To do this, we selected all the objects classified as galaxies i.e. non-AGN (note that this comparison sample includes both star-forming and passive galaxies) in the Stripe 82 database for which spectroscopic redshifts was determined and was matched in redshift and galaxy luminosity with the sample of QSO hosts (see F14 for details).  In order to build a control sample that match the characteristics of our multicolor sub-sample of 52 QSO we select all the galaxies with $z<0.3$ and that are well matched in terms of absolute magnitude of quasar host and redshift distribution. The control sample of galaxies turned out to have 83 objects with a mean redshift $<z>$=$0.24\pm0.05$ and average absolute magnitude of $<M_i>$ = $-22.52\pm0.68$, indistinguishable from that of the QSO hosts, see Figure \ref{fig:miz}.  

For all the objects in this comparison sample of inactive galaxies we obtained from the Stripe82 catalogs the magnitudes in all the five Sloan bands. All these magnitudes, as for the QSO host galaxies magnitudes were corrected for extinction using the SDSS data \citep{abazajian09} and k-corrected using the  package KCORRECT \citep{Blanton2007}.  

\section{Image analysis}
\label{Sect:analysis}
For the 52 QSO we retrieved the calibrated and combined  images in all colors ($u, g, r, i, z$) from SDSS Stripe 82 dataset  \citep{annis2011}. These final combined frames were obtained by co-adding selected fields that, in r-band, have a seeing (as derived from 2D gaussian fit of stars and provided by SDSS pipeline) better than 2 arcsecs, sky brightness $\mu$=19.5 mag/$arcsec^2$ and less than 0.2 mag of extinction; for this reason in all the 5 bands the final frames have the same number of exposures co-added. The images used have an average seeing, as given by 2D Gaussian fit of stars in the frame from SDSS, of 1.49$\pm$0.10 arcsecs in $u$, 1.41$\pm$0.06 arcsecs in $g$, 1.27$\pm$0.06 arcsecs in $r$, 1.2$\pm$0.07 arcsecs in $i$ and 1.25$\pm$0.07 in $z$.

In order to derive the properties of the galaxies hosting the QSO we performed a 2D fit of the
images of the QSO following the same procedure adopted for the analysis of the full sample 
\citep{falomo14}. Briefly we assume that the image of the QSO  is the superposition of two components. The nucleus in the center and the surrounding nebulosity (the host galaxy) . The first is described by the local Point Spread Function of the image while for the second component we assumed  a galaxy model described by a Sersic law convolved with the proper PSF. The analysis of these images was performed using the Astronomical
Image Decomposition Analysis package \citep[AIDA,][]{uslenghi08}.

The most critical aspect of the image decomposition is the determination of a suitable PSF. In
the case of SDSS images the field of view is large enough that there are always many stars in the
 co-added SDSS image containing the target to properly derive the PSF. 
 As noted in our previous work  \citep{falomo14} the PSF provided by SDSS pipeline, although it is computed from the stars in the frame that are close to the position of the target, it does not account properly for the shape of the PSF at radii larger than about 3 arcsec. The difference between the psField PSF and the true radial profile of stars was shown in Figure 4 of F14. The net effect of using psField PSF is that in all cases where the signal from the quasars extends more than 3 arcsec from the center of the image the decompositions in terms of point source plus a host galaxy may be  systematically biased. In these cases host galaxies are overestimated and in a number of cases true unresolved sources are confused with resolved objects. 
 
\begin{figure}
\centering
\includegraphics[width=0.95\columnwidth]{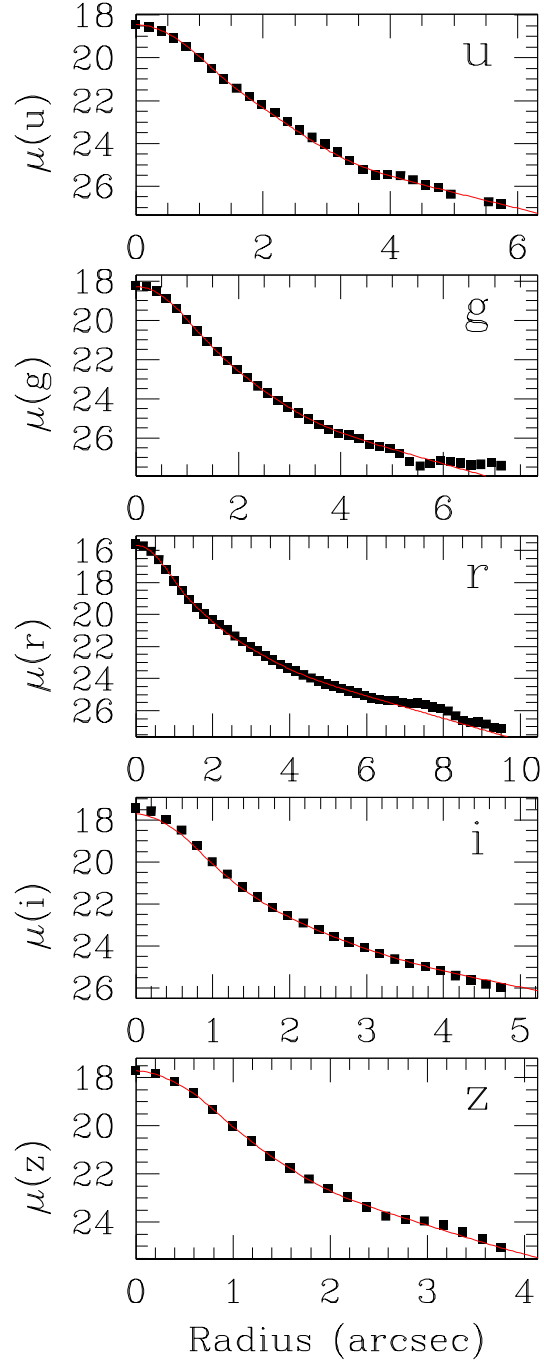}
\caption{Example (object N.19) of the PSF model (red solid line) compared with the azimutally averaged radial brightness profiles of some stars in the frame (squares).}
\label{fig:sixpsf}
\end{figure}

 To derive a suitable PSF for the targets in each filter we selected for each field a number of stars around the QSO  and computed a PSF model composed by the combination of 3 gaussians and one exponential functions. 
 Then we looked at the fit ($\chi^2$  and visual inspection of the average radial brightness profile) of the model for each PSF stars  (c.f. details in  F14) in order to remove possible bad stars (poor fit) and fit again all the "good" stars to produce the final PSF model (see an example, in all filters, in Figure \ref{fig:sixpsf}).

Using these PSF models we then fit all images of the QSO and for all objects we computed the fit of the image of the targets in all filters : $u, g,r,i,z$ (filter "i" was already available) and then evaluated if the object is resolved (as in the i band) or not in all the remaining filters. We note that, as expected, the nucleus/host ratio increases toward u-band, leading to more unresolved cases there than in red filters, in fact only $\sim$50\% of objects is found resolved in u.

The last step of the analysis is to fit each quasar with a 2 components model (point source plus a galaxy). 
Again we proceeded with the same recipes as in \citet{falomo14}. For the co-added images we assume a readout noise of 9.5 e$^-$ and an average gain of 3.8 e$^-$/ADU. The term for the statistical noise is given by  the coefficient $1/\sqrt{GAIN \times NEXP}$ that multiply the root square of the counts. For the residual pattern noise we assumed 2\% value. 
The final classification of the targets in each filter was based on the comparison of $\chi^2$ for the two  
fit (only psf and psf + galaxy) and further visual inspection of the fit. From this procedure we classified 
all objects as resolved or unresolved. Two QSO are unresolved in all 5 bands and are present in our list for completeness.
\begin{figure*}
\centering
\includegraphics[width=2.0\columnwidth]{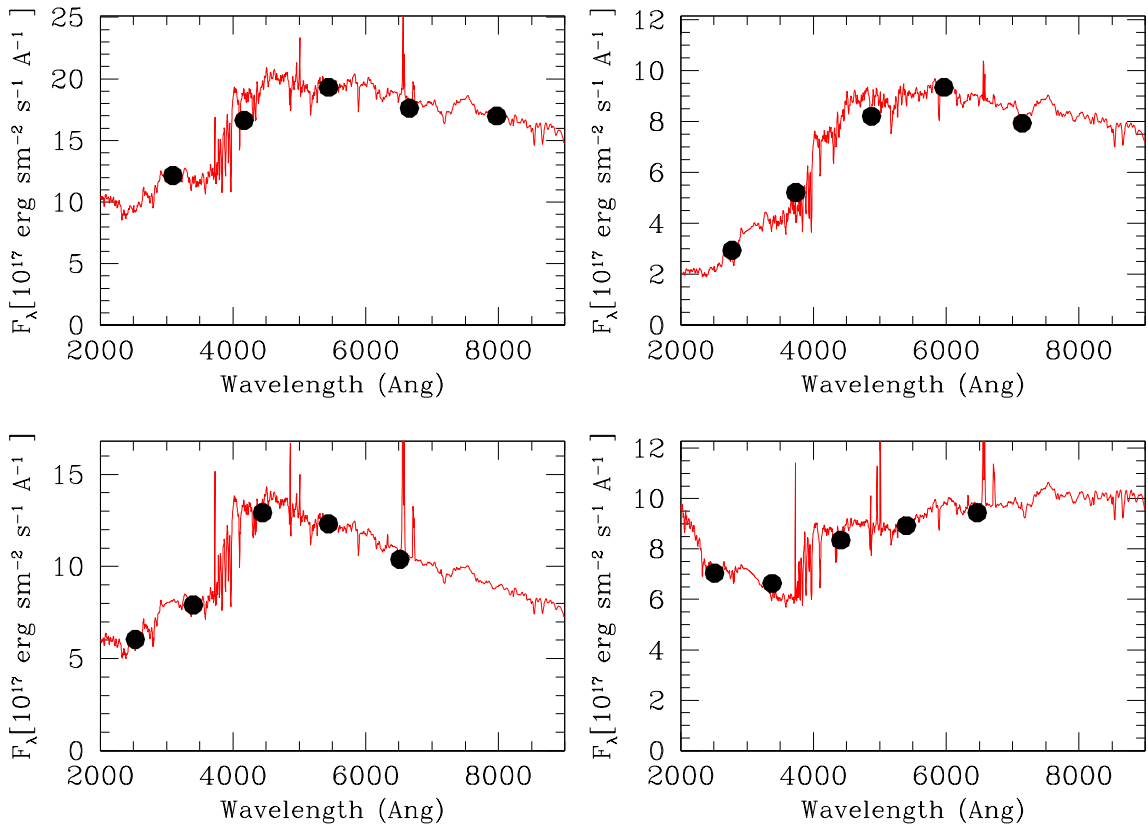}
\caption{Example of the SED fit (solid line) obtained from the KCORRECT tool for a number of QSO host galaxies  with different spectral flux distribution (filled circles). }
\label{fig:kcor}
\end{figure*}

To evaluate the errors on the fit parameters we performed a number of simulations of the targets and then 
compared the results of the analysis of the simulate data. For each object, we produced ten simulated images assuming the the best-fit parameters of the target but with different random noise and 
including somewhat different levels of the background (based on the uncertainty obtained from the original image). 
These simulated images were then processed using the same procedure adopted for the original one and the estimate of the error is derived from the comparison of the various best fit parameters.
From the distribution of the parameter values we assumed the semi-interquartile range as the error for the derived QSO parameters  (see table \ref{tab:sample}).

Because our main aim is to study the colors of the host galaxies when we compare colors we need to derive the K-correction for each filter.
We estimated the photometric K-corrections using the  package KCORRECT \citep{Blanton2007}, v.4.2, based on the fit of observed photometric points with   nonnegative linear combination of galaxy  spectral templates. 
We used the default set of 5 galaxy templates that are based on stellar population synthesis model of \cite{Bruzual2003}. 
K-corrections were evaluated at z=0 using the $ugriz$ magnitudes corrected for extinction. For half of the sources the fit was performed using all 5 band while for the other half only 4 filters were used. The quality of the fit is very good ( $\chi^2 < $1; see Figure \ref{fig:kcor}) for  90\% and 80\% of the galaxies, for the subsamples with 5 and 4 bands, respectively.

Similarly for the sample of inactive galaxies we obtain good fit for 85\% of the galaxies using in all cases 5 bands.



\section{Results}
\label{sect:hostgal}

\begin{figure}
\centering
\includegraphics[width=1.0\columnwidth]{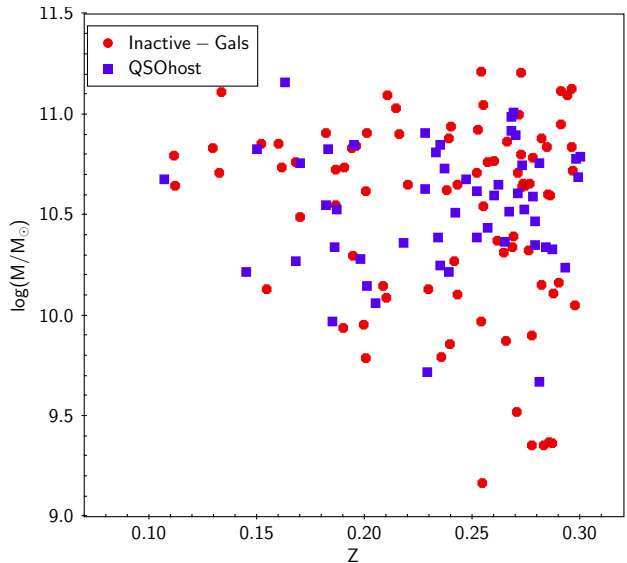}
\caption{Distribution of the mass of galaxies with respect to the redshift. The mass is estimated from the SED of the galaxies.}
\label{fig_mass}
\end{figure}

\begin{figure}
\centering
\includegraphics[width=1.0\columnwidth]{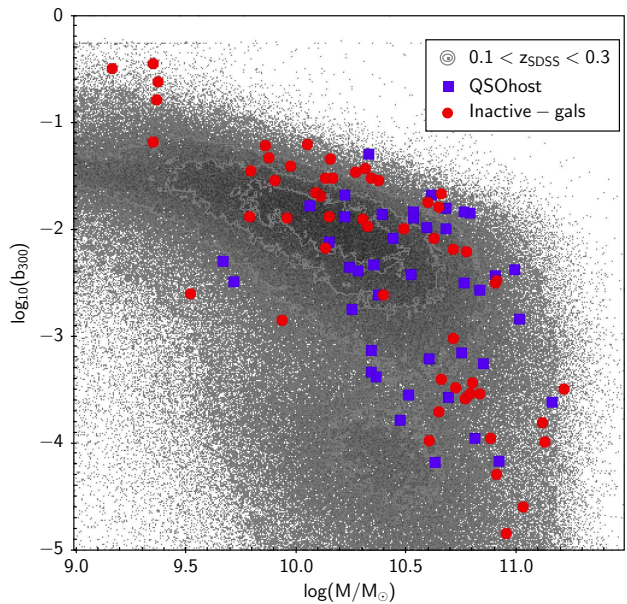}
\caption{Star formation, relative to average star-formation rate, that has occurred in the previous 300 Myr, with respect to the stellar mass (in solar units). Small grey dots and contours are the data for SDSS normal galaxies at $0.1<z<0.3$  \citep{Blanton2005}. }
\label{mas_age}
\end{figure}

In Figure \ref{fig:kcor} we show some example of the fit of the SED of QSO host galaxy using KCORRECT. 
From this fits it is also possible to derive some information of the age, stellar content and in particular for the presence of a young component. We also computed the stellar masses from the SED fitting and since KCORRECT assumes a cosmology with 
H$_0$ = 100 km s$^{-1}$Mpc$^{-1}$, the stellar masses were scaled to our cosmology as log (M*h$^{-2}$), where h = $H_0$/100 = 0.7. 
Also for the comparison sample of inactive galaxies we derived the stellar masses from the SED fitting and in figure \ref{fig_mass} we show the distribution of masses for both samples with respect to the redshift. The two samples show similar properties in particular, for our sample of resolved objects, we find an average mass of the QSO host galaxy of $<\mathcal{M}_*>=4.28\pm2.76\times10^{10}M_{\odot}$ and $<\mathcal{M}_*>=5.27\pm3.88\times10^{10}M_{\odot}$  for the comparison sample of normal galaxies. The SED fitting gives also information on the fraction of the total star formation, relative to average star-formation rate, that has occurred in the previous 300 Myr (b300) and 1Gyr (b1000). For QSO hosts and inactive galaxies we obtained both quantities, in figure \ref{mas_age} we  show the b300 parameter with respect to the stellar mass $M^*$. As comparison we plot also the distribution of normal galaxies from the SDSS spectroscopic sample at $0.1<z<0.3$ derived from the NYU Value-Added Galaxy catalog\footnote{sdss.physics.nyu.edu/vagc/} \citep{Blanton2005}. The plot indicate very similar properties, in particular if we consider that QSO hosts are mainly found at log(M/$M_{\odot})>$10 there are no differences among the two samples.

In Table \ref{tab:sample} we list the final apparent magnitudes of the host galaxies. When the galaxy is unresolved in one color no magnitude is reported. In column (1) we give the identification number from the sample of F14 in column (2) the SDSS identification, in column (3) the redshift and in column (4), (5), (6), (7) and (8) the magnitudes in u, g, r, i, z respectively. The apparent magnitudes are also corrected for extinction using the values given by the SDSS database \citep{abazajian09} and  k-corrected using the prescriptions given above. In Table \ref{tab:color_1} we list in column (3) the i-band absolute magnitude $M_i$ the colors in columns from (4) to (6), in column (7) the mass $\mathcal{M_*}$ (in $\mathcal{M}_{\odot}$) derived from the SED fitting and in column (8) the BH mass (in $\mathcal{M}_{\odot}$) from \citet{shen13}.

\subsection{Colors of QSO host galaxies}

 \begin{table}
 \caption{The average colors of QSO hosts and Inactive galaxies}
\begin{tabular}{|r|l|r|r|r|r|r|r|}
\hline
  \multicolumn{1}{c}{Sample} &
  \multicolumn{1}{c}{u-g} &
  \multicolumn{1}{c}{g-i } & 
  \multicolumn{1}{c}{r-i} &  
  \multicolumn{1}{c}{r-z} \\
\hline  
  QSO host & 1.32$\pm$0.25 &0.82$\pm$0.26 & 0.26$\pm$0.16 &  0.55$\pm$0.21 \\
  Galaxies & 1.51$\pm$0.54 &0.88$\pm$0.34 & 0.30$\pm$0.12 &  0.53$\pm$0.21 \\ 
\hline
\end{tabular}
\label{tab:colors}
\end{table}

We were able to resolve all the quasars in the sample in the filters $g,r,i$ and $z$ but one object in $g$ and another in $z$ (see Table \ref{tab:sample} for details). For filter $u$, due to the reduced contrast between starlight and  nuclear emission we can resolve the QSO only for 24 objects (46\%). In the following analysis therefore we consider the full sample in all colors but not $u$ and for color analysis including $u$ band for a subsample of QSO.

For the whole sample  the average color are : $<g-i>$ is 0.82$\pm$0.26 and  $<r-z>$ is 0.55$\pm$0.21. For the subsample of objects resolved also in $u$ the mean color $<u-g>$ is 1.32$\pm$0.25 while the average colors are: 
$<g-i>$ = 0.83$\pm$0.20 and  $<r-z>$ is 0.57$\pm$0.19 formally indistinguishable from the average values of the full sample (see Table \ref{tab:colors} )

\begin{figure}
\centering
\includegraphics[width=\columnwidth]{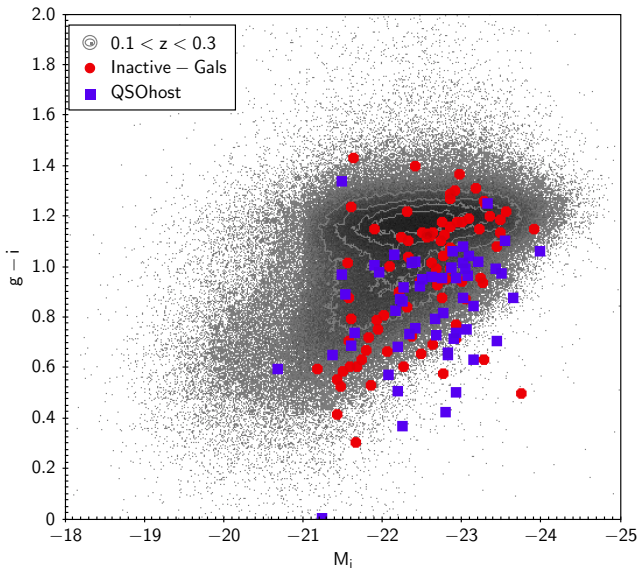}
\caption{Color-magnitude plot of the QSO host (blue squares), our sample of normal galaxies (red dots) superimposed to the distribution of the SDSS normal galaxies at $0.1<z<0.3$  (small grey dots and contours).}
\label{fig:comp_cmd}
\end{figure}



In figure \ref{fig:comp_cmd} we compare our data, for both QSO hosts and normal galaxies samples with the distribution of normal galaxies from the SDSS spectroscopic sample at $0.1<z<0.3$ derived from the MPA-JHU DR7 catalog\footnote{http://www.mpa-garching.mpg.de/SDSS/}. The two samples span the same range of values, however if we divide evenly our sample of QSOs it turns out that half of them have $M_i<$-22.5 and the other half $M_i>$-22.5. The average (g-i) color are: 0.82$\pm$0.26 and  0.75$\pm$0.32 respectively that compared with similar division for inactive galaxies (also divided at $M_i$=~~-22.5 produce roughly 50\% of subsamples) yields : 1.05$\pm$0.20 and 0.65$\pm$0.32. We note also a clear red sequence of galaxies, mainly populated by inactive galaxies. This contributes to the color difference, i.e., bluer color of QSO hosts compared to that of inactive galaxies, at the highest galaxy luminosities. This result for the luminous QSOs is in agreement with the suggestion \citep[see][]{kauff03,jahnke09,matsuoka14} that the most massive QSO host galaxies (those with $M_i \sim<$ -22) are bluer, and thus more star-forming, than inactive galaxies of similar luminosity. Both quasar hosts and inactive galaxies of similar mass/luminosity cover a wide range of colors ( 0.3 $< g-i < $  1.3) that are on average bluer than that of the bulk of normal galaxies. 

\begin{figure}
\centering
\includegraphics[width=\columnwidth]{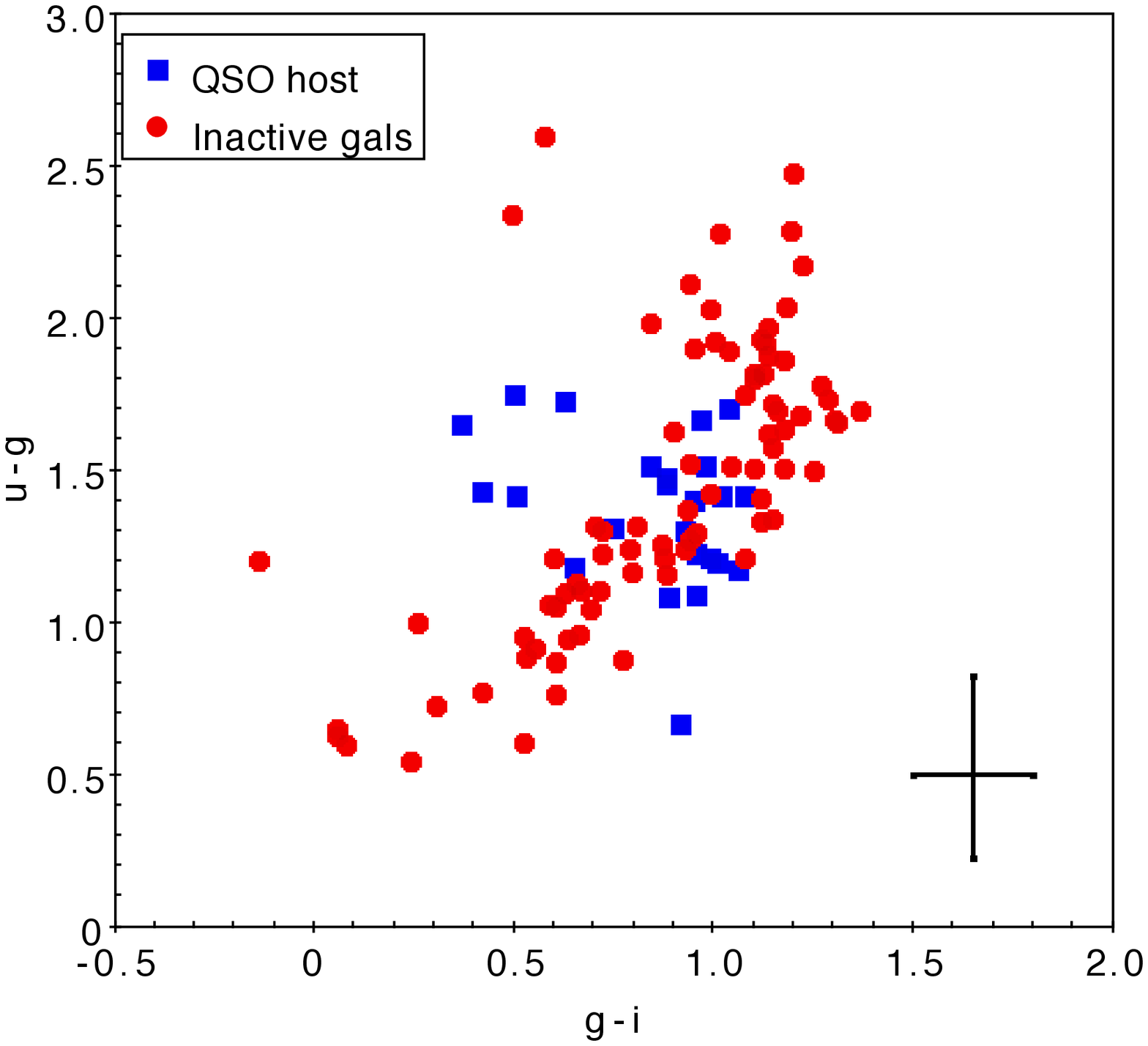}
\includegraphics[width=\columnwidth]{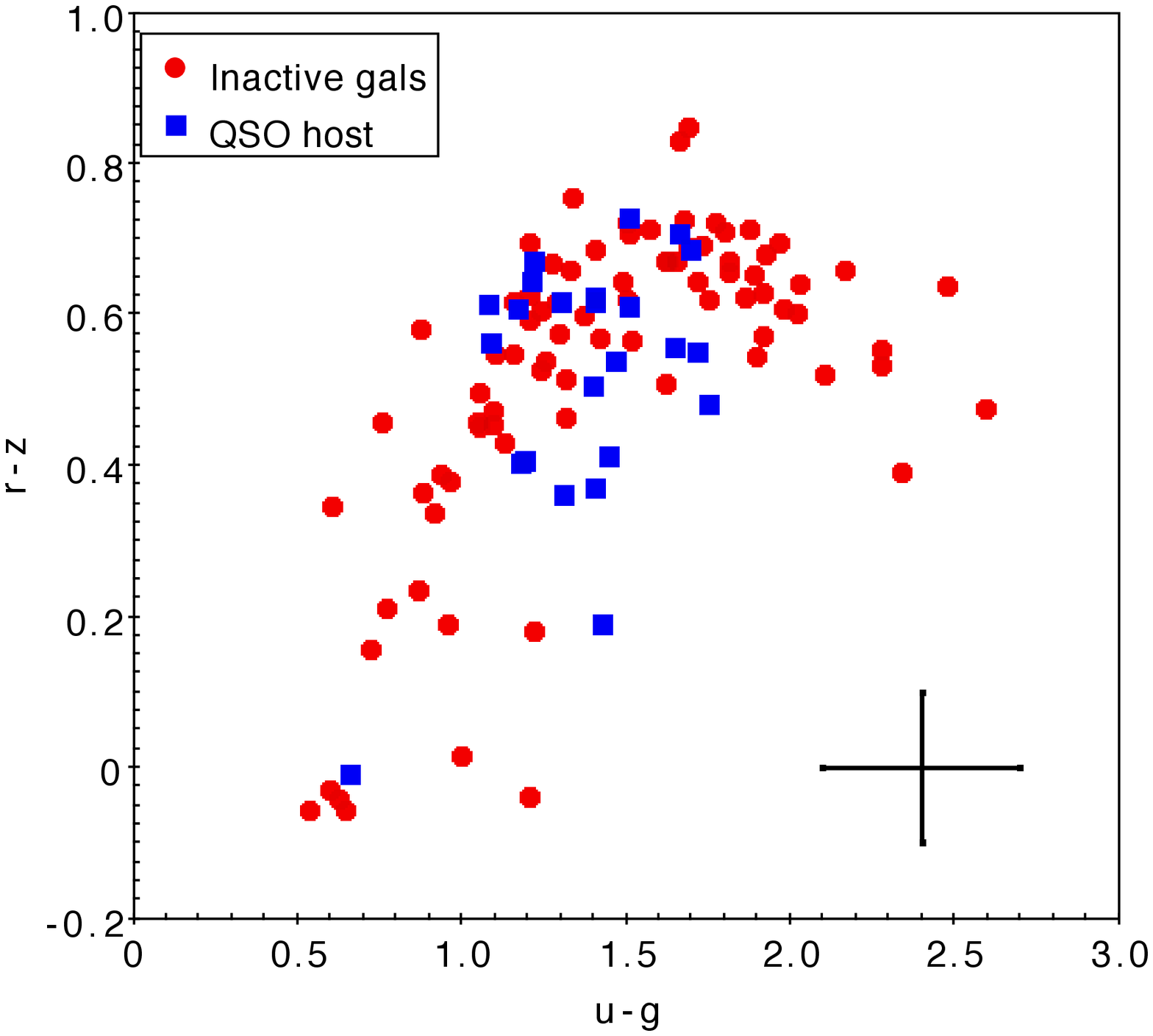}
\caption{Color-color diagrams of the quasar hosts compared to our sample of normal galaxies. In the lower right corner the average color error bars are plotted. }
\label{fig:col_gi}
\end{figure}

In figure \ref{fig:col_gi} we plot the comparison of the color-color diagrams (g-i vs. u-g) and (u-g vs. r-z) for quasar hosts and inactive galaxies. The average colors for the sample of inactive galaxies are: $<g-i>$=0.88$\pm$0.34, $<u-g>$=1.51$\pm$0.54 and $<r-z>$=0.53$\pm$0.21, formally indistinguishable, within the errors, from those of quasar hosts in our sample. However we note that the u-g colors of QSO hosts for objects resolved in u-band span a narrow range in color and have a smaller scatter of values than the inactive galaxies (see upper panel of fig. \ref{fig:col_gi}).  
Finally we note that the inactive galaxies exhibit a wider color range in both color diagrams, suggesting that the control sample is a mixture of massive red sequence galaxies and less massive star forming blue cloud galaxies.

\subsection{Close companion galaxies}

\begin{figure}
\includegraphics[width=1.2\columnwidth]{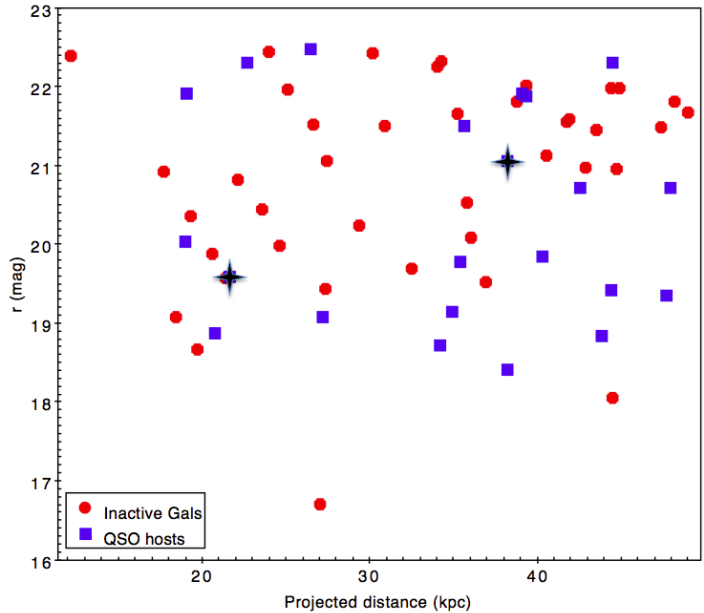}
\caption{The distribution of distance from the QSO (blue) and Inactive galaxies (red) versus r band apparent magnitude of the companion galaxies at projected distance less than 50 kpc. The crosses indicates the two close companions with the same redshift as the QSO.}
\label{fig:dist}
\end{figure}

For both samples of QSOs and inactive galaxies  we obtained from the Stripe82 catalogs the magnitudes in all bands of the objects classified as galaxies that are at a projected distance from the target less than 50 kpc (at the redshift of the QSO/Galaxy). 
To minimize the contamination of background objects, we consider as possible companions only the objects with $r<22.5$ i.e. four magnitudes fainter that the average r-band magnitude of our active or inactive galaxies. 
In Table \ref{tab:ncomp} we give the statistic of the number of companions found in both samples and in Figure \ref{fig:dist} we show the distribution of the projected distance (derived using the redshift of the targets) versus r-band apparent magnitude of the close companion for both QSOs and galaxies. 

\begin{figure}
\centering
\includegraphics[width=\columnwidth]{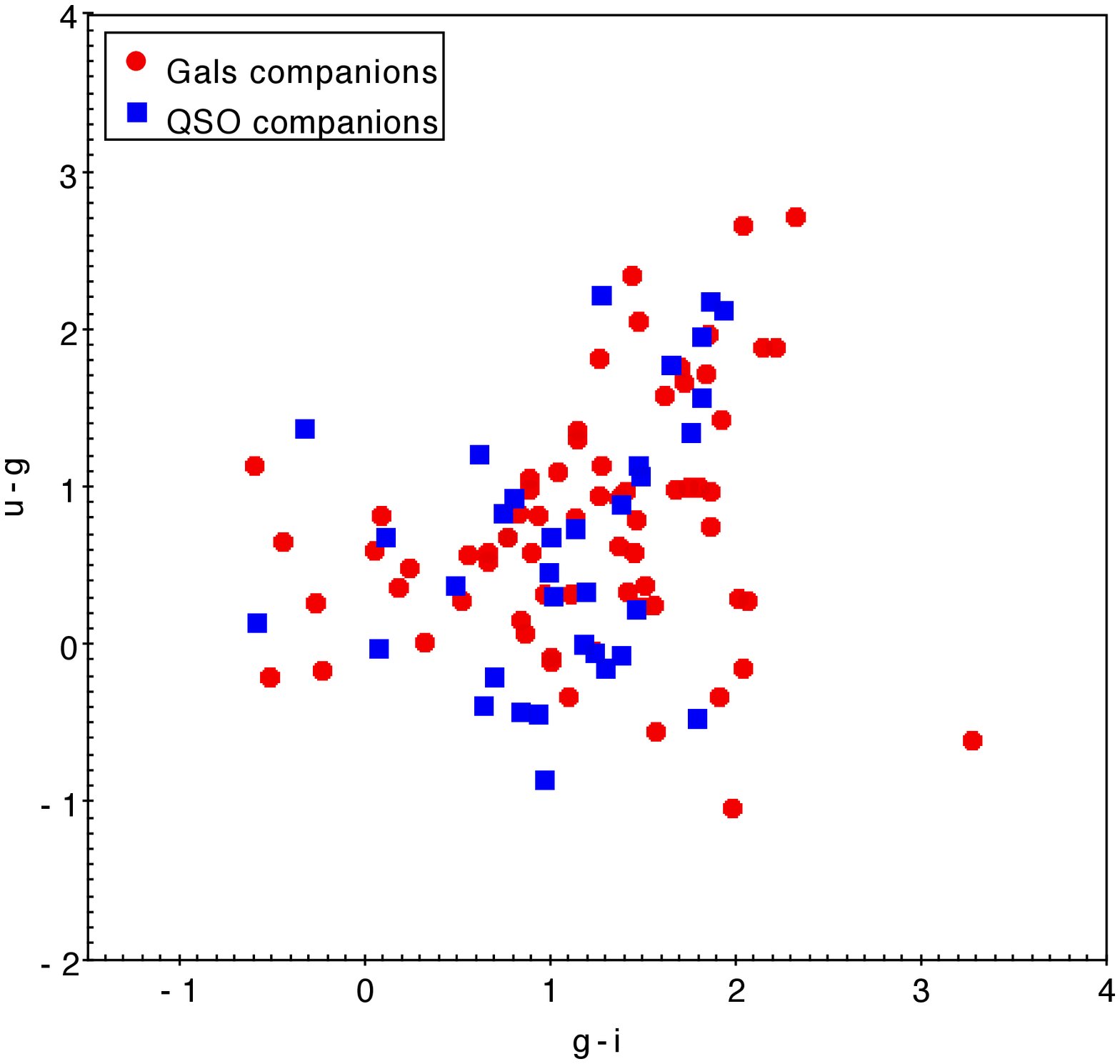}
\includegraphics[width=\columnwidth]{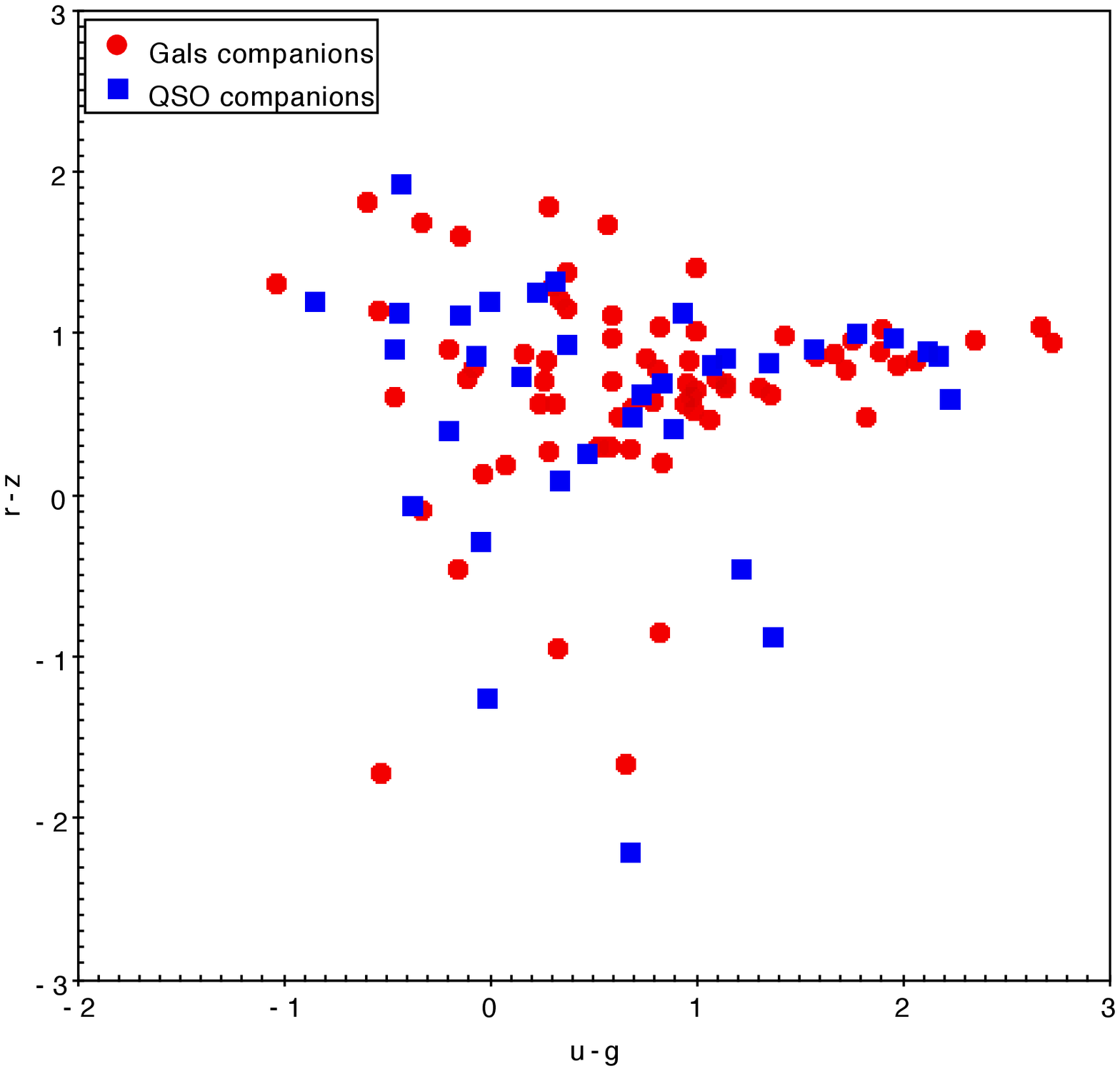}
\caption{The color-color diagrams of the close companions of bot QSO and galaxies. Top panel g-i vs. u-g and bottom panel u-g vs. r-z.}
\label{fig:dist1}
\end{figure}

The comparison of the statistics of the close companions for both QSO and matched inactive galaxies sample do not show significative differences (see Figure \ref{fig:dist}). There is the 35\% of QSO and 39\% of galaxies that do not have close companions and the percentages of objects in both samples that have 1 ($\sim$31\% and $\sim$29\%) or 2 ($\sim$25\% and $\sim$22\%) companions are very similar. Finally only $\sim$10\% of the remaining objects in both samples have more than 3 companions (see table \ref{tab:ncomp}). On average the  r magnitudes of the companions are very similar ($<r>=20.85\pm1.27$ for QSO and $<r>=20.36\pm1.32$ for the normal galaxies).  There is a suggestion that  bright (r$ <$20) companion galaxies are more frequent  in QSO ($\sim$ 50\%) than in inactive galaxies ($\sim$25\%). While this could be associated to the nuclear activity (past merging and/or interaction) a larger statistical sample is required  to reach a firm conclusion. The color-color plots for the companions for both QSO hosts and Galaxies are shown in figure \ref{fig:dist1}. The color of close companions of QSO and normal galaxies cover the same region in the explored bands. From our comparison thus there is no signature of bluer colors for the companions of active galaxies with respect to those of normal galaxies of similar mass.

For both QSO and inactive galaxies samples we also searched for spectroscopic data of the close companions. Only for 6 objects with at least one neighborhood we found the redshift of the companion. In four cases (\#61, \#92, \#95 and \#192) the companion is a foreground galaxy while for two objects (\#130 and \#200), the redshift is identical to that of the QSO. For the inactive galaxies we found a similar results. The redshift of companion was found for 5 objects and only in one case  it coincides with that of the target galaxy.  We point out that because of the lack of the redshift of the companions we can only compare statistically the frequency of the companions candidates between the two samples. If the majority of the companions be not associated with QSO/galaxy a possible difference of physically associated companions might be hidden.

\subsection{Comparison with previous studies}

 \begin{table}
 \caption{The statistics of close companions}
\begin{tabular}{|r|l|r|r|r|r|r|r|}
\hline
  \multicolumn{1}{c}{Sample} &
  \multicolumn{1}{c}{0 } & 
  \multicolumn{1}{c}{1} &  
  \multicolumn{1}{c}{2} &
  \multicolumn{1}{c}{$>$3} \\
\hline  
  QSO & 18 (35\%) & 16 (31\%) & 13 (25\%) & 5 (10\%) \\
  Galaxies & 36 (39\%) & 27 (29\%) & 20 (22\%) & 9 (10\%) \\ 
\hline
\end{tabular}
\label{tab:ncomp}
\end{table}

In this work we analyzed the color properties of the host galaxies and their immediate environment for  a homogeneous sample of low redshift  ($z < 0.3$)  QSO and compared their properties with those of a similar sample of inactive galaxies. 
As detailed in Sect. \ref{Sect:analysis} we are able to resolve the QSO in $g,r,i,$ and $z$ band for $>$95\% of the objects in the sample and for $\sim$ 50\% also in the $u$ band. This ensures that we are able to explore the color properties of the host galaxies of QSO with little incompleteness effects due to the increasing fraction of unresolved sources at higher redshift. 
Another reason to set a low limit (z = 0.3) to the redshift of the sample is that with this limit it is possible to extract a significant sample of normal (inactive) galaxies with similar absolute magnitude of the QSO host galaxies.

The color distribution of QSO hosts covers a region (see Fig \ref{fig:comp_cmd}) of bluer color with respect to the bulk of galaxy population (red sequence). The majority of the objects  are in the region between the star-forming and the quiescent galaxies \citep[see also][]{salim07}. Similar results were reported by \citet{kauff03} and \citet{jahnke04}  for a sample of low z AGN and also by M14 for their sample of QSO in Stripe82.  However, when QSO hosts are compared with a matched sample of inactive galaxies the average colors are found  very similar and a bluer color for the quasar hosts appears for the most luminous host galaxies (see Sect. 4.1)

The determination of the luminosity of the host galaxies of a sizable sample of quasars allows one to investigate the relationship between the luminosity of the host and the mass of their supermassive black holes (SMBH). The latter can be derived with the virial method using the width of broad emission lines and the continuum luminosity (see e.g. \citet{shen11,shen13}). Note that our sample of QSO is limited to z$ < $0.3 where cosmic evolution effects (or selection effects), as those discussed by \citet{SW} are negligible (or not detectable).
Based on the Stripe82 quasars M14 find a positive correlation (albeit with a quite large dispersion) between the  BH mass and the host galaxy luminosity and/or mass. From their comparison with the local relations derived by \citet{HR} (note that the comparison is with the erratum of M14 i.e. \citet{matsuoka14e}) for inactive galaxies
they conclude that quasar hosts are found to be under-massive for a given SMBH mass. 
Alternatively one could interpret this difference as higher mass of the BH for a given mass/luminosity of the galaxy.

This is an opposite result of that found by F14 for a similar sample of S82 QSOs at z $<$ 0.5 (see F14; Figures 12 and 14). Since the BH mass of QSO are obtained from the same source \citep{shen11} the difference should arise from the evaluation of the luminosity of the host galaxies.
The comparison is somewhat  complicated by the fact that M14  give absolute magnitudes in AB system and refer to $i$ band assuming objects at z = 0.3 while F14 transform the observed magnitudes into rest frame R filter (Vega magnitude) in order to be able to compare the host luminosities with previous results secured with HST for QSO of similar redshift. 
 
\begin{figure}
\centering
\includegraphics[width=\columnwidth]{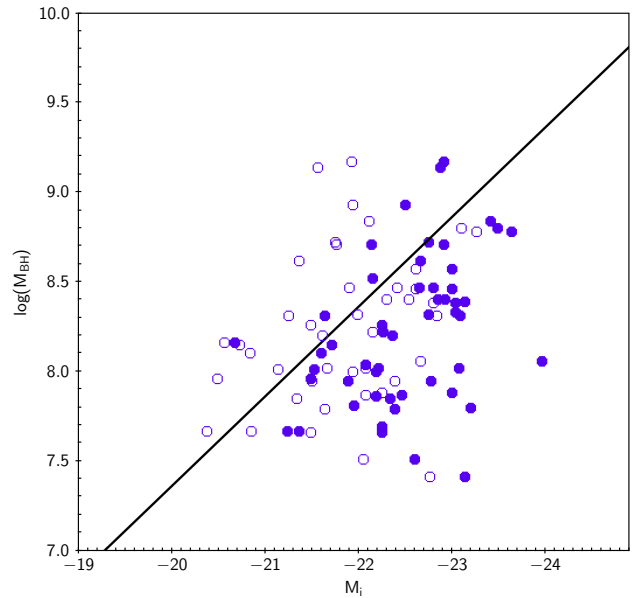}
\caption{Absolute magnitude (AB) in $i$ band of QSO host galaxies (filled blue circles) versus BH mass (in solar masses) for 52 resolved quasars at z $<$ 0.3.
The reference (black) solid line is the \citet{bettoni03} relation for local (inactive) galaxies for which black hole mass was measured.
The majority of the host galaxies lie below the local relation and are suggestive of the presence of a significant disc component not correlate with the central BH mass (see text for details). Open circles refers to the bulge component only.}
\label{fig:mbhmi}
\end{figure}

 In Figure \ref{fig:mbhmi} we report the relationship between M(BH) and host galaxy absolute magnitude $M_i$ compared with the local relation by \citet{bettoni03} that is in good agreement with those by \citet{MD} and \citet{Ferrarese} based on similar datasets.
 The local relation by \citep{bettoni03} that is calibrated on the R (Vega mag system) filter was transformed into $M_i$ taking into account both the different cosmology assumed and the filter transformations.
 Contrary to the claim of M14 we do not find a significant positive correlation between the two quantities and find that most of the objects are located below the local relation. This is consistent with our previous finding based on a much larger sample (see F14) and was interpreted as due to a significant disc component in the QSO host galaxy  \citep[see e.g.][]{decarli2012}.  At low BH masses a significant disk component has been also recently noted by \citet{Sanghvi14} and \citet{GS15}.  For all the objects in the sample we estimated the bulge to total galaxy luminosity based on morphological classification T of the host galaxy following the scheme in \citet{nair10}. We divided our objects in five morphological classes  (see F14 for details) and assigned a bulge/disk ratio ranging from 1 for T=-5 to 0.3 for T=2. It turns out that, when the disk component is removed, a more significant relationship is found between BH mass and the bulge component of the host galaxies (see figure \ref{fig:mbhmi}). 
      
 
\bigskip


\section{Summary and Conclusions}

We have investigated  the color properties of the host galaxies and their close environment from an  homogeneous dataset of 52  low redshift (z $<$ 0.3) quasars using the $u,g,r,i$ and $z$ SDSS images in the Stripe82 region.
The 2D analysis of the images allowed us to well resolve the 
quasar host for  almost all the objects in the sample in g, r, i and z filters and only for half in the u-band. The color properties together with the statistics of close companion galaxies of quasars are compared with those of a homogeneous sample of inactive galaxies at similar redshift and comparable luminosity.
The following  properties of quasar hosts are derived:

\begin{enumerate}

\item
The overall mean colors of the QSO host galaxy are indistinguishable from those of inactive galaxies of similar luminosity and redshift.  There is a suggestion that the most massive QSO hosts have bluer colors and show a lower star formation rate, in the last 300 Myr, than the control sample of inactive galaxies. 

 \item 
For about 60\% of the quasars we found companion galaxies at projected distance less than 50 kpc. However, the fraction of objects that have companions at the same redshift of the QSO  appears to be only  $\sim$10\%. Moreover the comparison with the companions of inactive galaxies indicates that very similar fractions of companions are present also in non active galaxies of same  luminosity/mass.

\item 
We do not found a significant correlation between the central BH mass and the total luminosity of the quasar hosts. This is contrary to previous claims (M14) based on similar dataset that quasar hosts are found to be under-massive for a given SMBH mass.  We found that host of quasars are more luminous than expected from the local $M_{BH}-M_{bulge}$ relation and interpret it as suggestive of a disc component that is not correlated with the BH mass.

\end{enumerate}
The comparison of color properties of the quasar host galaxies and of the galaxies  in the immediate environments with those of a similar sample of non active galaxies do not indicate any significant difference.  This further support  a scenario where the activation of the nucleus has negligible effects on the global structural and photometrical properties of the hosting galaxies. In particular the similarity of colors between active and inactive galaxies of similar mass indicate that also the stellar content of these galaxies is virtually unchanged by the presence of an active nucleus.


\section*{Acknowledgments}

We thank the referee for constructive comments that improved this paper.
Funding for the SDSS and SDSS-II has been provided by the Alfred P. Sloan Foundation, the Participating Institutions, the
National Science Foundation, the U.S. Department of Energy, the National Aeronautics and Space Administration, the Japanese
Monbukagakusho, the Max Planck Society, and the Higher Education Funding Council for England. The SDSS Web Site is
http://www.sdss.org/.

The SDSS is managed by the Astrophysical Research Consortium for the Participating Institutions. The Participating
Institutions are the American Museum of Natural History, Astrophysical Institute Potsdam, University of Basel, University
of Cambridge, Case Western Reserve University, University of Chicago, Drexel University, Fermilab, the Institute for
Advanced Study, the Japan Participation Group, Johns Hopkins University, the Joint Institute for Nuclear Astrophysics, the
Kavli Institute for Particle Astrophysics and Cosmology, the Korean Scientist Group, the Chinese Academy of Sciences (
LAMOST), Los Alamos National Laboratory, the Max-Planck-Institute for Astronomy (MPIA), the Max-Planck-Institute for
Astrophysics (MPA), New Mexico State University, Ohio State University, University of Pittsburgh, University of
Portsmouth, Princeton University, the United States Naval Observatory, and the University of Washington.

%

 \begin{table*}
 \caption{The full sample of resolved objects$^1$}
\begin{tabular}{|r|l|r|c|c|c|c|c|}
\hline
  \multicolumn{1}{|c|}{Nr} &
  \multicolumn{1}{c|}{SDSS} &
  \multicolumn{1}{c|}{z} &
  \multicolumn{1}{c|}{$m_u$(host)} &
  \multicolumn{1}{c|}{$m_g$(host)} &
  \multicolumn{1}{c|}{$m_r$(host)} &
  \multicolumn{1}{c|}{$m_i$(host)} &
  \multicolumn{1}{c|}{$m_z$(host)} \\
\hline
  18 & 205105.02-003302.7 & 0.3   & 20.50$\pm$0.92 & 18.83$\pm$0.24 & 18.55$\pm$0.12 & 17.86$\pm$0.26 & 17.85$\pm$0.29\\
  19 & 205212.28-002645.2 & 0.268 & 20.14$\pm$1.53 & 18.73$\pm$0.18 & 17.93$\pm$0.24 & 17.65$\pm$0.16 & 17.32$\pm$0.27\\
  21 & 205418.80+004915.9 & 0.228 & --    & 18.08$\pm$0.16 & 17.63$\pm$0.58 & 17.05$\pm$0.61 & 17.11$\pm$0.37 \\
  36 & 211234.88-005926.8 & 0.235 & 19.45$\pm$0.19 & 18.36$\pm$0.07 & 17.70$\pm$0.14 & 17.40$\pm$0.15 & 17.14$\pm$0.11\\
  40 & 211832.75+004500.8 & 0.233 & 19.50$\pm$0.79 & 18.32$\pm$0.18 & 17.70$\pm$0.09 & 17.30$\pm$0.15 & 17.29$\pm$0.17\\
  43 & 212203.82+001119.2 & 0.229 & --     & 20.18$\pm$0.72 & 19.56$\pm$0.54 & 19.58$\pm$0.57 & --  \\
  44 & 212556.26+004539.3 & 0.281 & --    & 19.54$\pm$0.40  & 19.18$\pm$0.21 & 19.53$\pm$0.16 & 18.95$\pm$0.34 \\
  51 & 213110.54-003537.0 & 0.145 & 19.61$\pm$0.99 & 18.53$\pm$0.10  & 17.92$\pm$0.06 & 17.63$\pm$0.06 & 17.31$\pm$0.21\\
  52 & 213245.24+000146.4 & 0.234 & --    & 19.44$\pm$0.31 & 18.72$\pm$0.12 & 18.43$\pm$0.12 & 17.95$\pm$0.18\\
  59 & 215408.71-002744.4 & 0.218 & --    & 19.05$\pm$0.25 & 18.50$\pm$0.37 & 18.00$\pm$0.15 & 17.79$\pm$0.08\\
  61 & 215516.13+003250.8 & 0.278 & 20.77$\pm$0.50  & 19.36$\pm$0.24 & 18.80$\pm$0.14 & 18.34$\pm$0.20 & 18.18$\pm$0.37\\
  62 & 215744.18+005303.6 & 0.267 & 20.41$\pm$0.72 & 19.11$\pm$0.09 & 18.47$\pm$0.08 & 18.18$\pm$0.08 & 17.85$\pm$0.16\\
  68 & 215949.01+001004.7 & 0.271 & --    & 18.82$\pm$0.25 & 18.37$\pm$0.38 & 18.03$\pm$0.27 & 17.73$\pm$0.39\\
  89 & 222315.11-002610.5 & 0.293 & --   & 20.73$\pm$0.43 & 19.96$\pm$0.09 & 19.39$\pm$0.35 & 19.20$\pm$0.59\\
  92 & 222632.66-005717.7 & 0.168 & 19.24$\pm$0.66 & 17.83$\pm$0.21 & 17.37$\pm$0.46 & 17.33$\pm$0.39 & 17.00$\pm$0.27\\
  95 & 222909.81+002527.3 & 0.228 & 19.90$\pm$1.00  & 18.21$\pm$0.21 & 17.50$\pm$0.09 & 17.17$\pm$0.26 & 16.81$\pm$0.19\\
  113 & 230007.27+001739.1 & 0.265 & --   & --    & 19.34$\pm$0.27 & 18.91$\pm$0.26 & 18.71$\pm$0.29\\
  127 & 231250.88+001719.0 & 0.257 & --   & 18.95$\pm$0.22 & 18.41$\pm$0.21 & 18.21$\pm$0.25 & 17.93$\pm$0.24\\
  129 & 231625.39-002225.4 & 0.298 & 20.13$\pm$0.48 & 18.42$\pm$0.18 & 18.06$\pm$0.10  & 17.78$\pm$0.09 & 17.51$\pm$0.13\\
  130 & 231711.79-003603.6 & 0.186 & --    & 18.77$\pm$0.25 & 18.10$\pm$0.16 & 17.80$\pm$0.19 & 17.51$\pm$0.11\\
  133 & 232259.98-005359.2 & 0.15  & 18.56$\pm$0.20  & 17.11$\pm$0.10  & 16.45$\pm$0.07 & 16.23$\pm$0.07 & 16.03$\pm$0.09 \\
  143 & 233816.42+005029.8 & 0.183 & 18.88$\pm$0.21 & 17.66$\pm$0.10  & 16.99$\pm$0.07 & 16.67$\pm$0.08 & 16.35$\pm$0.09 \\
  154 & 234932.77-003645.8 & 0.279 & --   & 18.30$\pm$0.14 & 18.56$\pm$0.24 & 18.50$\pm$0.24 & 18.09$\pm$0.39\\
  157 & 235251.87+003814.9 & 0.273 & --     & 18.88$\pm$0.10  & 18.16$\pm$0.12 & 17.82$\pm$0.09 & 17.31$\pm$0.10\\
  160 & 235441.54-000448.6 & 0.279 & 20.06$\pm$0.50  & 19.41$\pm$0.16 & 18.83$\pm$0.18 & 18.49$\pm$0.21 & 18.84$\pm$0.38\\
  161 & 235457.09+004219.9 & 0.27  & 19.66$\pm$0.24 & 18.16$\pm$0.14 & 17.60$\pm$0.10  & 17.18$\pm$0.10 & 16.87$\pm$0.21\\
  170 & 000557.23+002837.7 & 0.26  & --    & 19.23$\pm$0.26 & 18.46$\pm$0.13 & 18.21$\pm$0.28 & 17.42$\pm$0.16\\
  178 & 001346.52+003402.8 & 0.274 & 19.73$\pm$0.37 & 18.55$\pm$0.12 & 18.06$\pm$0.10 & 17.90$\pm$0.07 & 17.66$\pm$0.13\\
  189 & 002752.39+002615.6 & 0.205 & --    & 19.28$\pm$0.27 & 18.71$\pm$0.14 & 18.63$\pm$0.27 & 18.41$\pm$0.32\\
  192 & 002831.71-000413.3 & 0.252 & 19.84$\pm$1.05 & 18.09$\pm$0.79 & 17.77$\pm$0.49 & 17.59$\pm$0.64 & 17.29$\pm$0.29\\
  198 & 003711.00+002127.8 & 0.235 & 20.10$\pm$0.58 & 18.45$\pm$0.15 & 18.30$\pm$0.07 & 18.08$\pm$0.11 & 17.74$\pm$0.17\\
  199 & 003723.49+000812.5 & 0.252 & 19.57$\pm$0.30  & 18.14$\pm$0.09 & 17.80$\pm$0.10 & 17.72$\pm$0.15 & 17.61$\pm$0.12\\
  200 & 004032.10-001350.8 & 0.242 & --   & 19.07$\pm$0.39 & 18.38$\pm$0.37 & 18.24$\pm$0.32 & 17.92$\pm$0.23\\
  229 & 011254.91+000313.0 & 0.239 & --   & 19.47$\pm$0.24 & 18.88$\pm$0.09 & 18.73$\pm$0.14 & 18.42$\pm$0.15\\
  277 & 015521.69-004149.8 & 0.269 & 19.36$\pm$0.31 & 17.90$\pm$0.73 & 17.34$\pm$0.08 & 17.02$\pm$0.07 & 16.80$\pm$0.09\\
  288 & 015950.24+002340.8 & 0.163 & 17.71$\pm$0.37 & 16.53$\pm$0.24 & 15.86$\pm$0.10 & 15.47$\pm$0.21 & 15.26$\pm$0.08\\
  309 & 021359.79+004226.7 & 0.182 & 19.56$\pm$0.23 & 18.15$\pm$0.10  & 17.51$\pm$0.12 & 17.20$\pm$0.14 & 17.01$\pm$0.18\\
  325 & 023922.87-000119.5 & 0.262 & --   & 18.66$\pm$0.09 & 18.20$\pm$0.14 & 17.93$\pm$0.26 & 17.39$\pm$0.38\\
  327 & 024052.82-004110.9 & 0.247 & --  & 18.27$\pm$0.13 & 17.76$\pm$0.20 & 17.53$\pm$0.13 & 17.29$\pm$0.22\\
  332 & 024340.98-002601.2 & 0.268 & --    & 18.23$\pm$0.07 & 17.56$\pm$0.09 & 17.23$\pm$0.11 & 16.95$\pm$0.09 \\
  333 & 024508.67+003710.7 & 0.299 & --    & 19.06$\pm$0.18 & 18.35$\pm$0.10 & 18.07$\pm$0.13 & 17.59$\pm$0.39\\
  335 & 024601.25-005937.2 & 0.201 & --    & 19.03$\pm$0.16 & 18.57$\pm$0.21 & 18.34$\pm$0.21 & 18.03$\pm$0.29\\
  339 & 025007.02+002525.3 & 0.198 & --    & 19.39$\pm$0.38 & 18.70$\pm$0.40 & 18.42$\pm$0.18 & 18.12$\pm$0.29\\
  342 & 025334.57+000108.3 & 0.17  & --    & 17.75$\pm$0.16 & 17.07$\pm$0.11 & 16.79$\pm$0.09 & 16.50$\pm$0.11\\
  349 & 025938.15+004216.3 & 0.195 & 19.08$\pm$0.41 & 17.57$\pm$0.11 & 16.93$\pm$0.08 & 16.72$\pm$0.07 & 16.32$\pm$0.09\\
  358 & 030639.57+000343.1 & 0.107 & 18.01$\pm$0.37 & 16.79$\pm$0.08 & 16.15$\pm$0.08 & 15.83$\pm$0.06 & 15.48$\pm$0.12\\
  360 & 030731.58+001558.4 & 0.284 & --    & 19.42$\pm$0.19 & 18.67$\pm$0.19 & 18.55$\pm$0.12 & 18.70$\pm$0.35 \\
  367 & 031142.02-005918.9 & 0.281 & 19.79$\pm$0.36 & 18.47$\pm$0.11 & 18.03$\pm$0.12 & 17.72$\pm$0.09 & 17.67$\pm$0.17 \\
  375 & 032213.89+005513.4 & 0.185 & --    & 18.25$\pm$0.35 & 17.73$\pm$0.30 & 17.67$\pm$0.19 & 17.69$\pm$0.31 \\
  390 & 033156.88+002605.2 & 0.237 & --    & 18.42$\pm$0.11 & 17.60$\pm$0.09 & 17.60$\pm$0.14 & 17.33$\pm$0.10 \\
  402 & 033651.52-001024.7 & 0.187 & --    & 18.42$\pm$0.15 & 17.81$\pm$0.10 & 17.55$\pm$0.21 & 17.26$\pm$0.14\\
  411 & 034430.03-005842.7 & 0.287 & --    & 19.32$\pm$0.27 & 19.01$\pm$0.16 & 18.64$\pm$0.15 & 18.50$\pm$0.15\\
\hline\end{tabular}
\begin{list}{}
\item   {\bf $^{(1)}$ All the reported magnitudes have been k-correctet} \\
\end{list}
\label{tab:sample}
\end{table*}

 \begin{table*}
 \caption{The colors of resolved objects}
 \begin{tabular}{|r|r|r|r|r|r|r|r|}
\hline
  \multicolumn{1}{|c|}{Nr} &
  \multicolumn{1}{c|}{z} &
  \multicolumn{1}{c|}{$M_i$} &
  \multicolumn{1}{c|}{u-g} &
  \multicolumn{1}{c|}{g-i} &
  \multicolumn{1}{c|}{r-i} &
  \multicolumn{1}{c|}{log($\mathcal{M}_*$)} &
  \multicolumn{1}{c|}{log($M_{BH}$)} \\
\hline
  18 & 0.3 & -23.08 & 1.66 & 0.97 & 0.69 & 10.79 & 8.02\\
  19 & 0.268 & -23.01 & 1.41 & 1.08 & 0.29 & 10.92 & 8.57\\
  21 & 0.228 & -23.2 & -- & 1.02 & 0.57 & 10.63 & 7.8\\
  36 & 0.235 & -22.93 & 1.09 & 0.96 & 0.3 & 10.85 & 8.4\\
  40 & 0.233 & -23.01 & 1.19 & 1.01 & 0.39 & 10.81 & 8.46\\
  43 & 0.229 & -20.68 & -- & 0.59 & -0.03 & 9.72 & 8.16\\
  44 & 0.281 & -21.24 & -- & 0.01 & -0.35 & 9.67 & 7.67\\
  51 & 0.145 & -21.53 & 1.08 & 0.89 & 0.29 & 10.22 & 8.01\\
  52 & 0.234 & -21.89 & -- & 1.01 & 0.29 & 10.39 & 7.95\\
  59 & 0.218 & -22.15 & -- & 1.05 & 0.5 & 10.36 & 8.71\\
  61 & 0.278 & -22.4 & 1.41 & 1.02 & 0.45 & 10.59 & 7.79\\
  62 & 0.267 & -22.47 & 1.3 & 0.93 & 0.29 & 10.52 & 7.87\\
  68 & 0.271 & -22.66 & -- & 0.8 & 0.34 & 10.61 & 8.47\\
  89 & 0.293 & -21.49 & -- & 1.34 & 0.57 & 10.24 & 7.96\\
  92 & 0.168 & -22.19 & 1.41 & 0.51 & 0.05 & 10.27 & 7.86\\
  95 & 0.228 & -23.09 & 1.7 & 1.04 & 0.33 & 10.91 & 8.31\\
  113 & 0.265 & -21.72 & -- & -- & 0.43 & 10.37 & 8.15\\
  127 & 0.257 & -22.34 & -- & 0.73 & 0.2 & 10.44 & 7.85\\
  129 & 0.298 & -23.14 & 1.72 & 0.63 & 0.28 & 10.78 & 8.39\\
  130 & 0.186 & -21.96 & -- & 0.98 & 0.31 & 10.34 & 7.81\\
  133 & 0.15 & -23.01 & 1.45 & 0.88 & 0.21 & 10.83 & 7.88\\
  143 & 0.183 & -23.05 & 1.21 & 0.99 & 0.32 & 10.83 & 8.33\\
  154 & 0.279 & -22.25 & -- & -0.2 & 0.05 & 10.35 & 8.26\\
  157 & 0.273 & -22.88 & -- & 1.06 & 0.34 & 10.75 & 9.14\\
  160 & 0.279 & -22.27 & 0.66 & 0.92 & 0.34 & 10.47 & 8.22\\
  161 & 0.27 & -23.49 & 1.51 & 0.98 & 0.42 & 10.9 & 8.8\\
  170 & 0.26 & -22.37 & --  & 1.01 & 0.25 & 10.6 & 8.2\\
  178 & 0.274 & -22.81 & 1.18 & 0.65 & 0.16 & 10.53 & 8.47\\
  189 & 0.205 & -21.37 & -- & 0.65 & 0.09 & 10.06 & 7.67\\
  192 & 0.252 & -22.92 & 1.75 & 0.5 & 0.19 & 10.62 & 8.71\\
  198 & 0.235 & -22.25 & 1.65 & 0.37 & 0.22 & 10.25 & 7.66\\
  199 & 0.252 & -22.78 & 1.43 & 0.42 & 0.08 & 10.39 & 7.95\\
  200 & 0.242 & -22.16 & -- & 0.83 & 0.14 & 10.51 & 8.52\\
  229 & 0.239 & -21.64 & -- & 0.74 & 0.15 & 10.22 & 8.31\\
  277 & 0.269 & -23.65 & 1.47 & 0.88 & 0.32 & 11.01 & 8.78\\
  288 & 0.163 & -23.97 & 1.17 & 1.06 & 0.39 & 11.16 & 8.06\\
  309 & 0.182 & -22.5 & 1.4 & 0.95 & 0.31 & 10.55 & 8.93\\
  325 & 0.262 & -22.67 & -- & 0.73 & 0.27 & 10.65 & 8.62\\
  327 & 0.247 & -22.92 & -- & 0.74 & 0.23 & 10.68 & 9.17\\
  332 & 0.268 & -23.42 & -- & 0.99 & 0.33 & 10.99 & 8.84\\
  333 & 0.299 & -22.86 & -- & 1.0 & 0.28 & 10.69 & 8.4\\
  335 & 0.201 & -21.6 & -- & 0.69 & 0.23 & 10.15 & 8.1\\
  339 & 0.198 & -21.49 & -- & 0.97 & 0.28 & 10.28 & 7.96\\
  342 & 0.17 & -22.75 & -- & 0.96 & 0.28 & 10.76 & 8.32\\
  349 & 0.195 & -23.15 & 1.51 & 0.84 & 0.21 & 10.85 & 7.41\\
  358 & 0.107 & -22.61 & 1.22 & 0.96 & 0.31 & 10.68 & 7.51\\
  360 & 0.284 & -22.25 & -- & 0.86 & 0.12 & 10.34 & 7.69\\
  367 & 0.281 & -23.05 & 1.31 & 0.75 & 0.31 & 10.76 & 8.38\\
  375 & 0.185 & -22.08 & -- & 0.57 & 0.06 & 9.97 & 8.04\\
  390 & 0.237 & -22.75 & -- & 0.82 & 0.0 & 10.73 & 8.72\\
  402 & 0.187 & -22.22 & -- & 0.87 & 0.26 & 10.53 & 8.02\\
  411 & 0.287 & -22.19 & -- & 0.68 & 0.37 & 10.33 & 8.0\\
\hline\end{tabular}
\label{tab:color_1}
\end{table*}


\begin{thebibliography}{}
\bibitem[\protect\citeauthoryear{Abazajian et al.}{2009}]{abazajian09} Abazajian, K.~N., Adelman-McCarthy, J.~K., Ag{\"u}eros, M.~A., et al.\ 2009, ApJ, 182, 543
\bibitem[\protect\citeauthoryear{Annis et al.}{2014}]{annis2011}Annis J., et al., 2014, ApJ, 794, 120 
\bibitem[\protect\citeauthoryear{Bahcall et al.}{1997}]{bahcall97} Bahcall J.N., Kirhakos S., Saxe D.H., Schneider D.P., 1997, ApJ, 479, 642
\bibitem[\protect\citeauthoryear{Baldry et al.}{2004}]{baldry} Baldry, I. K., Glazebrook, K., Brinkmann, J., et al. 2004, ApJ, 600, 681
\bibitem[\protect\citeauthoryear{Bettoni et al.}{2003}]{bettoni03} Bettoni D., Falomo R., Fasano G., Govoni F., 2003, A\&A, 399, 869
\bibitem[\protect\citeauthoryear{Blanton et al.}{2005}]{Blanton2005} Blanton M.~R., et al., 2005, AJ, 129, 2562 
\bibitem[Blanton \& Roweis(2007)]{Blanton2007} Blanton, M.~R., \& Roweis, S.\ 2007, AJ, 133, 734
\bibitem[Bruzual \& Charlot(2003)]{Bruzual2003} Bruzual, G., \& Charlot, S.\ 2003, MNRAS, 344, 1000
\bibitem[\protect\citeauthoryear{Canalizo \& Stockton}{2013}]{Canalizo} Canalizo, G., \& Stockton, A.\ 2013, ApJ, 772, 132 
\bibitem[\protect\citeauthoryear{Cisternas et al.}{2011}]{cisternas11} Cisternas,M., Jahnke,K., Bongiorno,A., et al., 2011, ApJ, 741, L11 
\bibitem[\protect\citeauthoryear{Decarli et al.}{2010}]{decarli2010} Decarli R., Falomo R., Treves A., Labita M., Kotilainen J.~K., Scarpa R., 2010, MNRAS, 402, 2453 
\bibitem[\protect\citeauthoryear{Decarli et al.}{2012}]{decarli2012} Decarli R., Falomo R., Kotilainen J.~K., Hyv{\"o}nen T., Uslenghi M., Treves A., 2012, AdAst, 2012,  
\bibitem[\protect\citeauthoryear{Dunlop et al.}{2003}]{dunlop03} Dunlop J.S., McLure R.J., Kukula M.J., et al., 2003, MNRAS, 340, 1095
\bibitem[\protect\citeauthoryear{Falomo et al.}{2008}]{falomo08} Falomo R., Treves A., Kotilainen J.K., Scarpa R., Uslenghi M., 2008, ApJ, 673, 694
\bibitem[\protect\citeauthoryear{Falomo et al.}{2014}]{falomo14} Falomo R., Bettoni, D.,  Karhunen, K., Kotilainen J.K. and Uslenghi, M., 2014, MNRAS, 440, 476 (F14)
\bibitem[\protect\citeauthoryear{Ferrarese}{2002}]{Ferrarese} Ferrarese L., 2002, ApJ, 578, 90 
\bibitem[\protect\citeauthoryear{Frieman et al.}{2008}]{frieman08} Frieman, J. A. et al., 2008, AJ, 135, 338
\bibitem[\protect\citeauthoryear{Floyd et al.}{2004}]{floyd04} Floyd D.J.E., Kukula M.J., Dunlop J.S., et al., 2004, MNRAS, 355, 196
\bibitem[\protect\citeauthoryear{Gultekin et al.}{2009}]{gultekin09}Gultekin, K., Richstone, D. O., Gebhardt, K. et al. 2009, ApJ, 698,198
\bibitem[\protect\citeauthoryear{H{\"a}ring \& Rix}{2004}]{HR} H{\"a}ring N., Rix H.-W., 2004, ApJ, 604, L89 
\bibitem[\protect\citeauthoryear{Heckman \& Best}{2014}]{heckman14} Heckman T.~M., Best P.~N., 2014, ARA\&A, 52, 589 
\bibitem[\protect\citeauthoryear{Hyv{\"o}nen et al.}{2007}]{Hyv} Hyv{\"o}nen T., Kotilainen J.~K., Falomo R., {\"O}rndahl E., Pursimo T., 2007, A\&A, 476, 723
\bibitem[\protect\citeauthoryear{Husemann et al.}{2014}]{Husemann} Husemann B., Jahnke K., S{\'a}nchez S.~F., Wisotzki L., Nugroho D., Kupko D., Schramm M., 2014, MNRAS, 443, 755 
\bibitem[Jahnke et al.(2004)]{jahnke04} Jahnke, K., Kuhlbrodt, B., \& Wisotzki, L.\ 2004, MNRAS, 352, 399 
\bibitem[\protect\citeauthoryear{Jahnke et al.}{2009}]{jahnke09}Jahnke K., et al., 2009, ApJ, 706, L215 
\bibitem[\protect\citeauthoryear{Jordi, Grebel, \& Ammon}{2006}]{Jordi} Jordi K., Grebel E.~K., Ammon K., 2006, A\&A, 460, 339 
\bibitem[\protect\citeauthoryear{Graham \& Scott}{2015}]{GS15} Graham A.~W., Scott N., 2015, ApJ, 798, 54 
\bibitem[\protect\citeauthoryear{Karhunen et al.}{2014}]{karhunen14} Karhunen, K. Kotilainen, Falomo, R., Bettoni,D. Usleghi, M. 2015, in preparation
\bibitem[\protect\citeauthoryear{Karhunen et al.}{2015}]{karhunen15} Karhunen, K. Kotilainen, Falomo, R., Bettoni,D. Usleghi, M. 2014, MNRAS, 441, 1802
\bibitem[\protect\citeauthoryear{Kauffmann et al.}{2003}]{kauff03} Kauffmann G., et al., 2003, MNRAS, 346, 
1055 
\bibitem[\protect\citeauthoryear{Kocevski et al.}{2012}]{Kocevski} Kocevski, D.~D., Faber, S.~M., Mozena, M., et al. 2012, ApJ, 744, 148 
\bibitem[\protect\citeauthoryear{Kotilainen \& Falomo}{2004}]{kotilainen04} Kotilainen J.K., Falomo R.,2004, A\&A, 424, 107 
\bibitem[\protect\citeauthoryear{Kotilainen et al.}{2007}]{kotilainen07} Kotilainen J.K., Falomo R., Labita M., Treves A., \& Uslenghi M., 2007, ApJ 660 1039
\bibitem[\protect\citeauthoryear{Kotilainen et al.}{2009}]{kotilainen09} Kotilainen J.K., Falomo, R., Decarli, R., Treves, Uslenghi, M and Scarpa, R., ApJ 703 1663 
\bibitem[\protect\citeauthoryear{Kukula et al.}{2001}]{kukula01} Kukula M.J., Dunlop J.S., McLure R.J., et al., 2001, MNRAS, 326, 1533
\bibitem[\protect\citeauthoryear{Li et al.}{2008}]{li08} Li C., Kauffmann G., Heckman T.~M., White S.~D.~M., Jing Y.~P., 2008, MNRAS, 385, 1915 
\bibitem[\protect\citeauthoryear{Liu, Zakamska, \& Greene}{2014}]{Liu} Liu G., Zakamska N.~L., Greene J.~E., 2014, MNRAS, 442, 1303 
\bibitem[\protect\citeauthoryear{Matsuoka et al.}{2014}]{matsuoka14} Matsuoka Y., Strauss M.~A., Price T.~N., III, DiDonato M.~S., 2014, ApJ, 780, 162 (M14)
\bibitem[\protect\citeauthoryear{Matsuoka et al.}{2014a}]{matsuoka14e} Matsuoka Y., Strauss M.~A., Price T.~N., 
III, DiDonato M.~S., 2014a, ApJ, 789, 91 
\bibitem[\protect\citeauthoryear{Matsuoka et al.}{2015}]{matsuoka15} Matsuoka Y., et al., 2015, arXiv, 
arXiv:1506.07535 
\bibitem[\protect\citeauthoryear{McLeod \& Rieke}{1995}]{McRi} McLeod K.~K., Rieke, G. H., 1995, ApJLett., 454, L77 
\bibitem[\protect\citeauthoryear{McLure \& Dunlop}{2002}]{MD} McLure R.~J., Dunlop J.~S., 2002, MNRAS, 331, 795 
\bibitem[\protect\citeauthoryear{Miller \& Sheinis}{2003}]{Miller} Miller, J.~S., \& Sheinis, A.~I.\ 2003, ApJLett, 588, L9 
\bibitem[\protect\citeauthoryear{Nair \& Abraham}{2010}]{nair10} Nair, P.~B., \& Abraham, R.~G.\ 2010, ApJL, 714, L260
\bibitem[\protect\citeauthoryear{Nolan et al.}{2001}]{Nolan} Nolan, L.~A., Dunlop, J.~S., Kukula, M.~J., et al.\ 2001, MNRAS, 323, 308 
\bibitem[\protect\citeauthoryear{Osterbrock}{1991}]{Ost} Osterbrock D.~E., 1991, RPPh, 54, 579
\bibitem[\protect\citeauthoryear{Pagani et al.}{2003}]{pagani}Pagani, C., Falomo, R., \& Treves, A. 2003, ApJ, 596, 830
\bibitem[\protect\citeauthoryear{Ridgway et al.}{2001}]{ridgway01} Ridgway S., Heckman T., Calzetti D., Lehnert M., 2001, ApJ, 550, 122
\bibitem[\protect\citeauthoryear{Salim et al.}{2007}]{salim07} Salim S., et al., 2007, ApJS, 173, 267 
\bibitem[\protect\citeauthoryear{Sanghvi et al.}{2014}]{Sanghvi14} Sanghvi J., Kotilainen J.~K., Falomo R., 
Decarli R., Karhunen K., Uslenghi M., 2014, MNRAS, 445, 1261 
\bibitem[\protect\citeauthoryear{Schmidt \& Green}{1983}]{SG} Schmidt M., Green R.~F., 1983, ApJ, 269, 352 
\bibitem[\protect\citeauthoryear{Schulze \& Wisotzki}{2014}]{SW} Schulze A., Wisotzki L., 2014, MNRAS, 438, 3422 
\bibitem[\protect\citeauthoryear{Shen et al.}{2011}]{shen11} Shen, Y., Richards, G.T., Strauss, M.A., et al. 2011, ApJSS, 194, 45
\bibitem[\protect\citeauthoryear{Shen et al.}{2013}]{shen13} Shen Y., et al., 2013, ApJ, 778, 98 
\bibitem[\protect\citeauthoryear{Schawinski et al.}{2011}]{Schawinski} Schawinski, K., Treister, E., Urry, C.~M., et al.\ 2011, ApJLett, 727, LL31 
\bibitem[\protect\citeauthoryear{Schneider et al.}{2010}]{schneider2010} Schneider, D.~P., Richards, G.~T., Hall, P.~B., et al.\ 2010, AJ, 139, 2360
\bibitem[\protect\citeauthoryear{Silverman et al.}{2008}]{Silverman} Silverman, J. D., Mainieri, V., Lehmer, B. D., et al. 2008, ApJ, 675, 1025 
\bibitem[\protect\citeauthoryear{Treister et al.}{2009}]{Treister} Treister, E., Virani, S., Gawiser, E., et al. 2009, ApJ, 693, 1713 
\bibitem[\protect\citeauthoryear{Uslenghi \& Falomo}{2008}]{uslenghi08} Uslenghi M. \& Falomo, R. 2008, Proc. of the 6th International Workshop on Data Analysis in Astronomy, Erice, World Scientific Pub., New Jersey, p.303
\bibitem[\protect\citeauthoryear{Weiner et al.}{2005}]{weiner} Weiner, B. J., Benjamin J., Phillips, A. C., et al. 2005, ApJ, 620, 595
\end{thebibliography}
\end{document}